\documentstyle[psbox]{WORKSHOP}
\def\lap{\hbox{{\lower -2.5pt\hbox{$<$}}\hskip -8pt\raise -2.5pt\hbox{$\sim$}}}
\def\gap{\hbox{{\lower -2.5pt\hbox{$>$}}\hskip -8pt\raise -2.5pt\hbox{$\sim$}}}

\def\BB#1\par{
  \noindent\hangindent=20pt
  { #1 }
  \par
}

\def\Mesz{M\'esz\'aros\ }
\def\Pacz{Paczy\'nski\ }
\newcount\fignumber
\def\newfig{\global\advance\fignumber by 1}
\def\fignam#1#2{\xdef#1{\the\fignumber}\newfig}
\def\fignam#1#2{\xdef#1{\the\fignumber}\hfil\break [REMOVE FIG NAME #2:]\newfig}

\newcount\eqnumber
\def\neweq{{\the\eqnumber}\global\advance\eqnumber by 1}
\def\eqnam#1#2{\xdef#1{\the\eqnumber}}

\def\lasteq{\advance\eqnumber by -1 {\the\eqnumber}\advance
     \eqnumber by 1}

\begin{document}

\title{
Relativistic Shells and Gamma-Ray Bursts
}

\author{
E. E. Fenimore and M. C. Sumner
\\ 
%
MS D436, Los Alamos National Laboratory, Los Alamos NM USA \\
{\it E-mail: efenimore@lanl.gov}
}
\abst{
In many models of Gamma-Ray Bursts (GRBs) relativistic shells are
 responsible for the overall envelope of emission.  We use
kinematics and symmetry to  calculate
the time history and spectral evolution expected from a relativistic
shell including effects from intrinsic variations in the shell's
intensity and spectra.  We find that the decay phase of an envelope is produced
by  photons delayed by the shell's curvature.  These delayed
photons are produced by regions that are off-axis such that the
spectra evolve according to a universal function
($\propto T^{-1}$) regardless of
intrinsic variations in the rest frame of the shell.
  We compare
our predictions to the overall envelope of emission of GRBs. 
The observed spectra
evolve faster. Intrinsic variations cannot make
the spectra evolve that fast, which adds strength to the  ``shell symmetry''
problem: models, in particular, the external shock model, that
involve  relativistic shells must either  
confine the material  to narrow pencil beams, be very
inefficient, or break the local spherical symmetry so that
the shell acts like a parallel slab.  In the case of the internal shock
models involving winds (i.e., central engines), it will probably be
easier to break the local spherical symmetry, but then one must
postulate nearly continuous energy generation at $10^{51}$ erg s$^{-1}$
lasting up to hundreds of seconds at the central site.
 }

\kword{gamma-ray bursts, relativistic shells}

\maketitle

\section{Introduction}

Gamma-ray Bursts (GRBs) have defied explanation since their discovery
25 years ago (Klebesadel, Strong, \& Olsen 1973).  The mystery was
deepened by the observation by the Burst and Transient Source
Experiment (BATSE) that GRBs are isotropic, yet there are fewer dim
bursts than expected relative to the bright bursts (Meegan et al. 1992).
 Two conflicting
explanations have arisen. In the ``cosmological'' explanation (see,
e.g., \Pacz 1995),
bursts originate at redshifts the order of $\sim 1$ and literally
flood the entire universe with gamma-rays.
 In the alternative ``galactic'' explanation
(see e.g., Lamb 1995), the bursts originate on high-velocity neutron
stars which have traveled sufficient distances ($\sim 100$ kpc) from
the disk of the Galaxy that we appear to be at the center of their
spatial distribution.


In both explanations,  the lack of apparent photon-photon attenuation
of high energy photons ($\sim 1 $ Gev) implies substantial bulk
relativistic motion.  The bulk Lorentz factor, $\Gamma$, must be the
on order of $10^2$ to $10^3$ for the cosmological explanation, and $10$
to $10^2$ for the galactic explanation.

 Two major scenarios involving relativistic shells have been developed.  In the
``external'' shock models (see, \Mesz \& Rees 1993,
Piran, Shemi, \& Narayan 1993), a  relativistic shell
is generated by a single, short-duration explosion at 
a central site.  The shell coasts
in a gamma-ray quiet phase for a time $t_0$.  Eventually, the shell
becomes gamma-ray active, perhaps due to shocks formed as the shell
sweeps up the ISM. (Hence, the name ``external'' shock.)
  Variations in the
shell explain the subpeaks observed in GRBs, but the overall envelope
of emission should follow that for a single expanding shell.
 There is a single, very thin, relativistic
shell, and its properties predict the overall envelope of emission of
the GRB.

Instead of an impulsive, single injection of energy, ``internal'' shock
models (Rees \& \Mesz 1994) feature a ``central engine'' 
that releases energy on timescales
comparable to the observed duration of the GRBs (up to hundreds of
seconds). In some models, the central engine
produces a
relativistic wind which develops instabilities that generate gamma-ray
producing shocks. In others, it  produces a series
of relativistic shells, and the fast shells collide with the slow shells
resulting in a shock.  In either case, the emitting region need not
be a single, thin shell as in the external models.  The duration of the
GRB comes
from the activity of the central engine (perhaps hundreds of seconds) 
and the individual peaks come
from inhomogeneities or instabilities in that activity.

\section{Expected Evolution from a Relativistic Shell}

In Fenimore, Madras, and Nayakshin (1996), we calculated the time
 history
 and spectral evolution expected from a relativistic shell under
the assumption of ``local spherical symmetry'' (LSS). By local, we mean that
only regions lying within angles the order of $\Gamma^{-1}$ contribute.
By spherical, we emphasize that the curvature of the shell within
$\Gamma^{-1}$ is the dominant feature that produces the
envelope.
By symmetry, we emphasize that the shell is seen head-on, rather than
edge-on, as in many scenarios involving jets.  Each of these assumptions
is justified in Fenimore, Madras, and Nayakshin (1996).
 
 We denote the rest
frame time of the detector as $t$ and the arrival time as $T$. 
While the shell is gamma-ray active,
the emitting surface keeps up with
the emitted gamma-rays so they arrive at the detector over a much
shorter period of time than they were emitted.
 If the
material directly on axis emits for time $t$, the photons arrive on a
time scale $T = (c-v)t/c \approx t/(2\Gamma^2)$, where $v$ is the bulk
velocity, $\Gamma = (1-\beta^2)^{-1/2}$, and $\beta = v/c$.
Thus, in the case of the external shock model,
the energy is released at the central site on a time scale of less
than a second, and the shell expands for $\sim 10^7$ s, emitting for a
 time scale somewhat shorter than $10^7$ s, but the photons
arrive within $\sim 10$ s (if $\Gamma = 10^3$).

 Let $P(\theta,\phi,R)$ give the rate of gamma-ray
production on the shell as a function of spherical coordinates. 
Individual bursts might have temporal structure
caused by variations in $P$ as a function of $(\theta,\phi)$; however,
 on average,
all off-axis regions should behave like the on-axis regions.
Since we are considering the average properties of bursts in this paper,
we assume
$P(\theta,\phi,R)$ is independent of $(\theta,\phi)$.
It is  a single, thin  shell, so $R = vt$ such that $P(\theta,\phi,R) = P(t)$.
The choice of a thin shell is justified by the observations. 
 If $\Delta T$ is the
typical time scale for variability in a GRB with duration $T_{\rm
dur}$, then the thickness relative to the size of the shell is $\Delta
r_{\parallel}/r_0 \sim \Delta T(2\Gamma^2 T_{\rm dur})^{-1}$ (see
Fenimore, Madras, and Nayakshin 1996).

\subsection{Temporal Structure from Emission at a Single Time}

Assume the shell
expands for a time ($t_e$) in a photon quiet phase and then emits for
a short period of time
 (i.e., $P(t) = P_0\delta(t-t_e)$). 
 In terms of arrival time, the
{\it on-axis} emission will arrive at  $T_e = t_e/(2\Gamma^2)$ and the
off-axis emission will arrive later with an expected shape
\eqnam{\VSINGLE}{VSINGLE}
\def\Vdelta{V_{\delta}(T,T_e)}
$$
\Vdelta  =  0~~~~~~~~~~~~~~~~~~~~~~~~~~~~~~~~~~~~~~~~~~{\rm if}~T<T_e
\eqno(\neweq a)
$$
$$
{}~~~~~~~~= \psi_1
T_e~\bigg({T \over T_e}\bigg)^{-\alpha-1} 
{}~~~~~~~~~~~~~~~~~~~~{\rm if}~T>T_e
\eqno(\lasteq b)
$$
where we have assumed that the rest frame photon-number
spectrum is isotropic with a power law with index $-\alpha$, and
$\psi_1$ is a constant.  Note in Fenimore, Madras, \& Nayakshin, 
this was incorrectly dervied as a power law with $-\alpha-2$. 
This envelope is similar to a ``FRED'' 
(fast rise, exponential decay)
where  the slow, power law decay depends only on
the time that the shell expands before it emits ($T_e$).
 The decay phase is due to
photons delayed by the curvature.

 GRB spectra
can often be fit by the so-called ``Band'' model (Band et al. 1993)
which consists of two power laws (indexes $\alpha_1, \alpha_2$) and a
connecting energy. If $E_c^{\prime}$ is the connecting energy in the
rest frame of the shell, the observed
connecting energy is boosted, $E_c = B_{\delta}(T,T_e) E_c^{\prime}$.
The delayed photons are  boosted less than at the peak because
they originate from regions moving at the angle $\theta$ relative to
the on-axis regions (see Katz, 1994).
 The boost as
a function of arrival time is (see
eq.~5 in Fenimore, Madras, Nayakshin 1996)
\eqnam{\BOOST}{BOOST}
$$
B_\delta(T,T_e) =\Lambda^{-1} = \Gamma(1-\beta\cos\theta)=
2\Gamma (T/T_e)^{-1}  \eqno(\neweq)
$$
This is the most important equation in this paper since it establishes
a one-to-one relationship between the angle at which photons originate
and the time at which they 
 {\it arrive} at
the detector if $T_e$ can be determined.
It is from this equation that we determine how the
observed spectra should evolve relative to how the time history evolves.

\subsection{Temporal Structure from Emission from a Range of Times}

More complex envelopes can be found from weighted sums of
$V_\delta(T,T_e)$.
  For example, the shell might emit for a range of times starting
at $t_0$.
We model  $P(T_e,T_0)$ to be non-zero  from $t_0$ 
to $t_{\rm max}$ and, we assume
$P(T_e,T_0)$ can be approximated as
\eqnam{\PTT}{PTT}
$$
P(T_e,T_0)= P_0(T_e/T_0)^{-\eta}.
\eqno(\neweq)
$$
 In terms of arrival time, the
{\it on-axis} emission will arrive between  $T_0 = t_0/(2\Gamma^2)$ and
$T_{\rm max} = t_{\rm max}/(2\Gamma^2)$.
 The
expected envelope, $V(T)$, is (see eq. 11 in Fenimore, Madras, \&
Nayakshin 1996):
\eqnam{\FUNENVELOP}{FUNENVELOP}
$$
V(T)
 = \int_{T_{0}}^{T}
V_{\delta}(T,T_e)P(T_e,T_0) \,dT_e
\eqno(\neweq)
$$
Due to the curvature of the shell, off-axis photons will be delayed, and  most 
emission will arrive later:
\eqnam{\ENVELOPE}{ENVELOPE}
$$
V(T) = 0  ~~~~~~~~~~~~~~~~~~~~~~~~~~~~~~~~~~~~~~~~~~~~~~~~~~{\rm if}~~T < T_{0}
\eqno(\neweq a)
$$
$$
{}~~~~~ =
{\psi_1 P_0 \over \omega T_0^{-\eta}} 
{T^{\omega} - T_{0}^{\omega} \over T^{\alpha+1}}
{}~~~~~~~~~~~~~~~~{\rm if}~T_{0} < T < T_{\rm max}
\eqno(\lasteq b)
$$
$$
{}~~~~ =
{\psi_1 P_0 \over \omega T_0^{-\eta}}
{T_{\rm max}^{\omega} - T_{0}^{\omega} \over  T^{\alpha+1}}
{}~~~~~~~~~~~~~~~~~{\rm if}~T > T_{\rm max}
\eqno(\lasteq c)
$$
where 
\eqnam{\POWERINDEX}{POWERINDEX}
$$
\omega=\alpha+3-\eta     \eqno(\neweq)
$$
Here, as a simplification to show the nature of the envelope,
we have used a power law spectral shape with a number index of
$\alpha$  ($\sim 1.5$). The deviations from the Band model are not
 important and $\alpha \sim (\alpha_1+\alpha_2)/2$.

The envelope in equation (\ENVELOPE) is also similar to a ``FRED''
where the  rise depends mostly on the duration of the photon
active phase
($T_{\rm max}-T_0$) and the slow, power law decay depends mostly on
the final overall size of the shell
($T_{\rm max}$).

The spectrum of the shell might change while it is emitting. 
 We model the possible change in $E_c^{\prime}$ as
\eqnam{\EC}{EC}
$$
E_c^{\prime}(T_e,T_0) = E_0^{\prime} (T_e/T_0)^{-\nu}
\eqno(\neweq)
$$
The observed connecting energy as a function of arrival time is
\eqnam{\AVEBOOST}{AVEBOOST}
$$
E_c(T)
 ={
 \int_{T_{0}}^{T}
V_{\delta}(T,T_e) P(T_e,T_0) E_c^{\prime}(T_e,T_0) B_{\delta}(T,T_e) \,dT_e
\over V(T)}
\eqno(\neweq)
$$
such that
\eqnam{\AVEBT}{AVEBT}
$$
E_c(T) =
{2\Gamma E_0^{\prime} \over T_0^{-\nu}}
 {\omega \over \omega^{\prime}}
{T^{\omega^{\prime}} - T_{0}^{\omega^{\prime}} \over
T^{\omega} - T_{0}^{\omega} }
T^{-1}
{}~~~~~~{\rm if}~T_{0} < T < T_{\rm max}
\eqno(\neweq a)
$$
$${}~~~~~~~
={2\Gamma E_0^{\prime} \over T_0^{-\nu}}
 {\omega \over \omega^{\prime}}
{T_{\rm max}^{\omega^{\prime}} - T_{0}^{\omega^{\prime}} \over
T_{\rm max}^{\omega} - T_{0}^{\omega} }
T^{-1}
{}~~~~~~~~~{\rm if}~T > T_{\rm max}
\eqno(\lasteq b)
$$
where
\eqnam{\OMEGAB}{OMEGAB}
$$
\omega^{\prime} = \omega -\nu +1  \eqno(\neweq)
$$

\section{Comparison to Observations}

In the external shock models, there is a single relativistic shell
which should provide 
the overall GRB emission envelope.  From the temporal structure of
GRBs, it is clear that some process is required to break the local spherical
symmetry and produce a chaotic time history rather than a smooth
profile as given by equation (\ENVELOPE).  However, on average, GRB time
histories should conform to the envelope if a single relativistic
shell is responsible.  Figure 1a is the average time history of 35
bright long GRBs (from Fenimore, 1997b).  It was generated by scaling
the measured duration to a standard length
which we call $T_{100}$ and then adding the 35 bursts together.  
($T_{100}$ is an average of $T_{50}/0.5$
and $T_{90}/0.9$)  Adding together in this fashion is
very reasonable for the external shock models.  Each burst appears to
have a different duration in arrival time because arrival time is
compressed by $\Gamma^{-2}$.
Normalizing each burst to a standard duration is equivalent to
 normalizing to a standard $\Gamma$. 

The average profile is very similar to a FRED, which is remarkable
because only 3 or 4 of the 35 bursts are clearly FRED-like.  It is
difficult to detect the FRED-like envelope in individual bursts
because the shell must be patchy in order to make a long complex burst
(Fenimore, Madras, \& Nayakshin 1996).
 However, the envelope becomes evident when many bursts are averaged
together. 

The dotted line in Figure 1a is the best fit of equation (\ENVELOPE)
to the average profile.  Besides an overall scale factor and $\alpha$,
there are only two free parameters: $T_0$ and $T_{\rm max}$. Note we
have assumed that $\Gamma$ is constant. From the fit in
 Figure 1a, $T_0 = 0.78
T_{100}$ and $T_{\rm max} = 0.85 T_{100}$.
We have shifted the $T$ scale such that zero is at the time (relative
to $T_{100}$) when the central explosion occurred.
 Thus, if a relativistic
shell is responsible for the shape in Figure 1a, we learn that the
gamma-ray quiet phase for the shell is $\sim 1.5 \Gamma^2 T_{100}$ and
the gamma-ray active phase is short: $0.07 \Gamma^2 T_{100}$.

Figure 1a implies that a relativistic shell can explain the average
envelope of emission of GRBs.
However, the single relativistic shell paradigm also predicts how the
spectra should evolve.  
The dotted curve in Figure 1b uses   the fit parameters from Figure
1a in equation (\AVEBT).  Since there are no additional free
parameters, the spectral evolution is entirely determined by the fit to
the time history.
We have averaged the 16 channel BATSE MER data in the same manner as
the averaging in Figure 1a.  The MER data were then summed into five
samples of duration  $0.15
T_{100}$ starting at $T = 1.0 T_{100}$.  The resulting spectra were
fit with  the ``Band'' spectral shape.
We determined $E_c$ in a manner similar to how Liang and Kargatis (1996) 
determined spectral evolution in GRB pulses.  We
first fit the entire time history to determine an average $\alpha_1$
and $\alpha_2$.  The five spectra were analyzed with these average power law
indexes such that $E_c$ was the only free
parameter.  In
Figure 1b, we have plotted the ratio of $E_c$ for each sample 
 to $E_c$ of the first sample. The first sample
was chosen to be in the decay phase of the envelope because the
spectral evolution (eq. [\AVEBT]) is effectively independent of the
intrinsic variation during the decay phase. The average GRB
envelope has spectra that evolve somwhat faster than  a
relativistic shell.   

%
%
\begin{figure*}[t]
\centering
\psbox[xsize=0.60#1,ysize=0.40#1]
{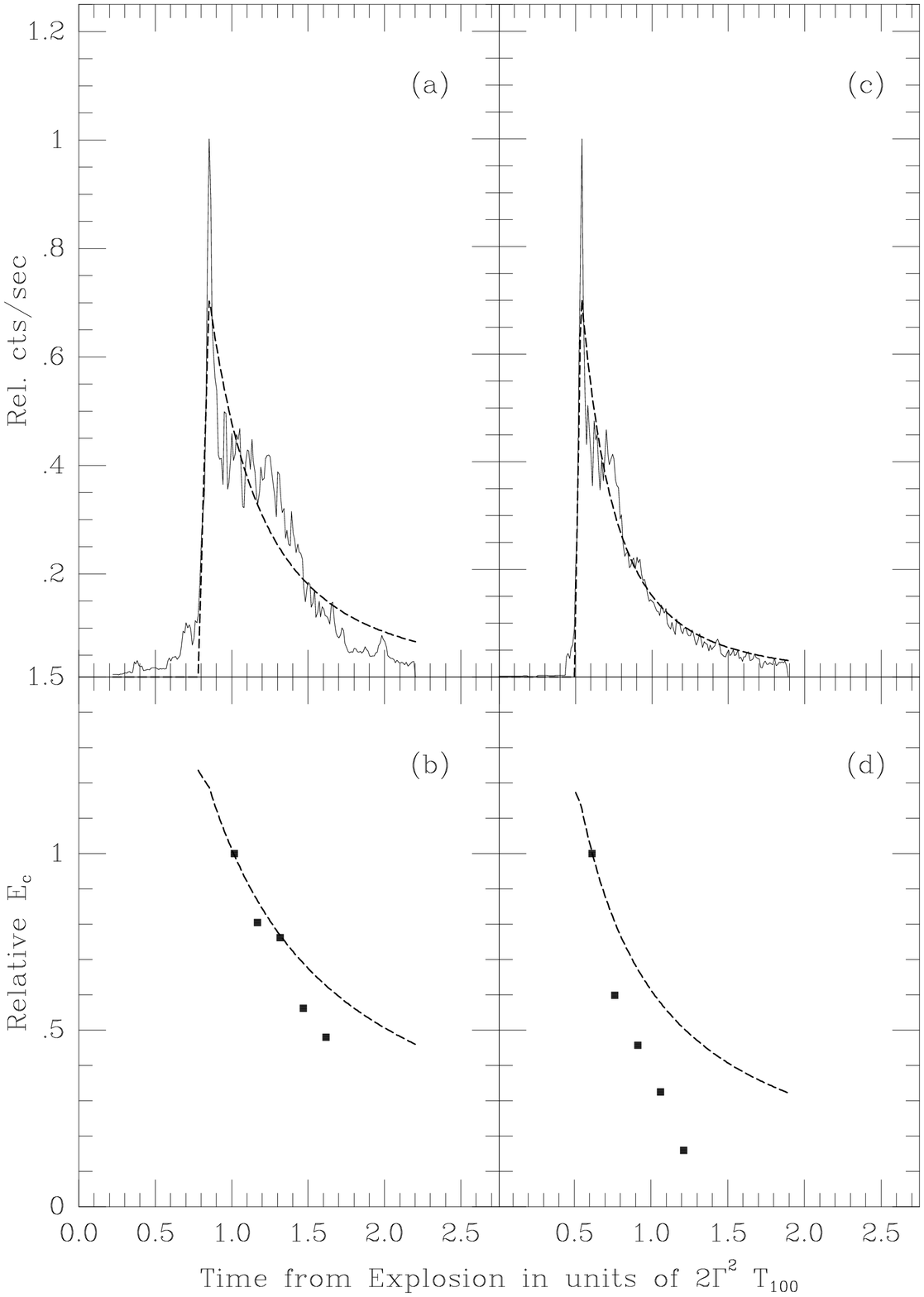}
\medskip
\caption{  
Comparisons  of time evolution expected from a single relativistic shell 
to GRB observations.  In the
``external'' shock model, the overall envelope of emission is related
to the properties of a single relativistic shell.
\hfill\break
({\it a}) Fits of a single relativistic shell to the
 average
 envelope of 35 bright, long BATSE bursts.
  The temporal scale of
 each individual burst was scaled to a standard
 duration called $T_{100}$. The dash line is the best
 fit solution based on equation
 (\ENVELOPE). 
The zero on the time scale
represents when the central explosion occurred
 and it is set
by the fit to equation (\ENVELOPE).
  The shell expanded in
 a gamma-ray quiet phase that lasted $\sim 0.78 T_{100}$ and then
 produced gamma-rays until $\sim 0.85 T_{100}$.
\hfill\break
({\it b}) Fit to the spectral evolution of the 35 bursts based on the
 BATSE MER data.  The dash line is
 the  expected relative characteristic energy in the spectrum based 
on the best fit
 parameters from ({\it a}) and equation (\AVEBT).  The
 solid squares are the observations.
\hfill\break
  ({\it c})Fits of a single relativistic shell to the
 average
 envelope of 6 FRED-like BATSE bursts. 
\hfill\break
({\it d}) The dash line is the expected  evolution based on
the fit in ({\it c}) and equation (\AVEBT). The squares are the 
observed spectral evolution
based on the BATSE MER data for the 6 FRED-like bursts.
\hfill\break
 The spectra of GRBs
 evolve faster than allowed  by the single relativistic shell
 that is utilized in the ``external'' shock models. 
The fits used in this figure are not substantially affected 
by intrinsic variation
in the shell's intensity or spectra. However, variations in the bulk
Lorentz factor might allow for a consistent fit of time history and spectra. 
}
\end{figure*}

It could be argued that the averaging of the 35 bursts is uncertain
because each burst is chaotic, and it is difficult  to properly
identify how to scale each one. Since FRED-like bursts have clear
envelopes and are the shape expected from a relativistic shell, they
should be easy to scale properly. 
In Figure 1c we averaged six FRED-like BATSE bursts,  stretched by
their durations to a standard time.  The decay phase fits better,
probably because the chaotic nature of the long complex bursts only
allows a rough determination of how much stretching is needed before
averaging. For the FRED-like bursts, the time of expansion from the
central site is somewhat smaller, $T_0 \sim 0.53 T_{100}$, and the rise
is very rapid, implying that $P(T_e,T_0)$ is non-zero only for a short
range of times.  In Figure 1d we show the expected evolution of the
connecting energy (dash curve) and the observed evolution (squares).
The conflict between the observations and the predictions from a single
relativistic shell is even stronger than observed with the 35 long
bright bursts.

%
%
\begin{figure*}[t]
\centering
\psbox[xsize=0.45#1,ysize=0.45#1]
{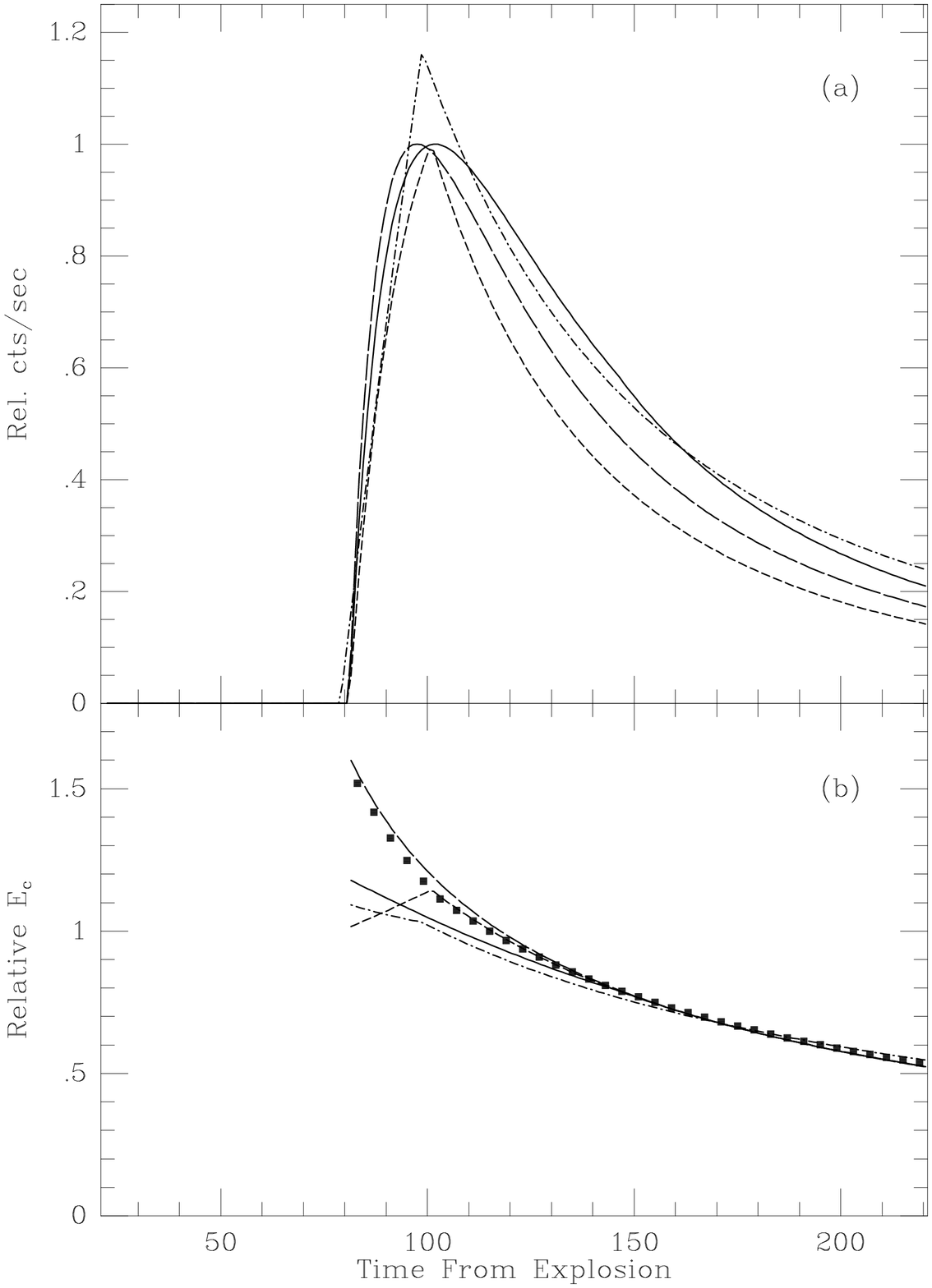}
\caption{
Variations in the observed time history and spectra of a relativistic
shell due to intrinsic variations in the rest frame of the shell.
\hfill\break
({\it a}) Time histories from a range of parameters in equation
(\ENVELOPE). The key parameters are the range of times the shell emits
in its rest frame ($T_0$ to $T_{\rm max}$) and how the emission
evolves with time ($\propto T^{-\eta}$). The solid, dot-short-dash,
short-dash, and long-dash curves correspond to ($T_0, T_{\rm max},
\eta$) = (80, 145, 9), (80, 100, 0), (80, 100, 5), and 
(80, $\infty$, 12), respectively. The intrinsic variation mostly
affects the rise portion of the envelope.  The decay phase is due to
off-axis emission which arrives later due to the curvature of the
shell. As a result, it is an average over the intrinsic variations and
follows a universal shape that is due to the
curvature ($\propto T^{-\alpha-1}$).
\hfill\break
({\it b}) Spectral evolution due to the intrinsic variation in the
shell. The parameters of the curves in ({\it a}) were used in equation
(\AVEBT) to give the evolution of a characteristic energy in the
spectrum. Here, $E_c$ is the energy that connects the two power laws in
the Band model. An additional parameter gives the intrinsic variation
of $E_c$ in the rest frame ($E_c^{\prime} \propto T^{-\nu}$). The four curves
used $\nu$ = 0, 0, -2, and +2, respectively. We fit the Band model to
simulated MER data using the parameters of the long-dash curve. The
resulting $E_c$'s are the squares.
Thus, if one averages a series of Band spectra, the $E_c$ of the resulting
spectrum is similar to the average 
of the individual $E_c$'s; there are no systematic effects
due to the use of the Band model.
\hfill\break
The intrinsic variation mostly affects the spectrum in the rise portion of the 
envelope.  The decay phase has spectra that are shifted, averages of
the spectrum during the rise.  The shift is the Lorentz factor
corresponding to the location of the off-axis emission responsible for
the decay portion.  The Lorentz factor varies as $T^{-1}$, so the
spectral evolution is a universal function ($\propto T^{-1}$) that 
is independent of the
intrinsic variation.
 }
\end{figure*}

\section{Intrinsic Spectral Variations}

In Figure 1, we assumed no intrinsic variations in the emission as a function
of time, that is, $\nu=\eta=0$ such 
that $P(T)=P_0$ and $E_c^{\prime} = E_0^{\prime}$.
In this section we show that intrinsic variations cannot provide
softening as rapid as seen in Figure 1.
 In Figure 2a, we use
a range of parameters in equation (\ENVELOPE) to demonstrate
the range of possibilities.  (All curves have $T_0 = 80$.)
The solid curve uses an intermediate
value of $T_{\rm max}$ (=145), and a value of $\eta$ of about
9 was used to obtain a fast rise and decay phase commensurate with the observed
characteristics of GRBs. The dot-short-dash curve was found as in
Figure 1 by fitting to the simulated data from the solid curve with
$\eta = 0$.  The resulting $T_{\rm max}$ is 98.  The short-dash curve
used $T_{\rm max}$ = 101 and  $\eta$ = 5.  Finally, the
long-dash curve uses $T_{\rm max} = \infty$ and $\eta = 12$. In all
cases, the decay phase is approximately  independent of the parameters used
 and scales roughly as $(T/T_0)^{-\alpha-1}$.

Figure 2b shows  the expected variation in $E_c$ based on
the parameters used for the curves in Figure 2a.  For the solid curve
and the dot-short-dash curve, we used no intrinsic variation in $E_c^{\prime}$
(i.e., $\nu = 0$).  For the short-dash and long-dash curves, we used
$\nu = -2$ and +2, respectively.  During the regions where
$P(T_e,T_0)$ is non-zero, the observed $E_c$ can be influenced by  intrinsic
variation in $E_c^{\prime}$.  However, in the decay phase, $E_c$ is
almost independent of $\nu$ and varies
as $(T/T_0)^{-1}$ (eq. \AVEBT, Fig. 2b), whereas the
observations (Fig. 1b, d) 
vary faster.

One assumption that we have implicitly 
made is that a weighted sum of Band spectra gives the
appearance of having a value of $E_c$ that is similar to the average
of the
values of individual $E_c$.
We tested this assumption by producing simulated spectra using the
parameters of the long-dash curve in Figure 2.
Simulated spectra  were produced for each time bin.
We processed the simulated spectra in the same manner as the MER data and
the results are shown as the squares in Figure 2.
In the decay phase, they closely follow the expected relationship.

\section{Discussion}

 The only 
approximations that we have made are that $P$ and $E_c^{\prime}$ are
approximately power laws (eqs. [\PTT] and [\EC]).  This allowed
analytic solutions which show the nature of equations (\FUNENVELOP)
and (\AVEBOOST).  These equations use only kinematic considerations to
demonstrate that the decay phase is influenced only by the curvature
of the shell.
During the decay phase, the delayed off-axis photons are strongly beamed,
and they are Lorentz-shifted out of the bandpass of the instrument.
The combination of
these two effects causes the time histories to decay as
$\sim \Lambda^{-\alpha-1} \sim
(T/T_0)^{-\alpha-1}$.
In contrast, the characteristic energy of the spectra varies as
$\sim
\Lambda^{-1} \sim (T/T_0)^{-1}$.
For example, by the time $T=2T_0$, the time history will have returned almost
to its baseline, while the characteristic energy will have changed only
by a factor of 2.  These relative rates of change are inconsistent with 
the observed overall envelope of GRB evolution.
 Indeed, it is well known that the characteristic
energy of  GRB spectra can change by factors the order of 4 
throughout a burst.

The fundamental reason that the spectra evolve as $T^{-1}$ can be seen in
equations (\AVEBOOST) and (\AVEBT). Notice that the only dependency on $T$ in
equation (\AVEBOOST) comes from $B_{\delta}(T,T_e)$ and
$V_{\delta}(T,T_e)$. To first order, the dependency on
$V_{\delta}(T,T_e)$ is cancelled by the $V(T)$ in the denominator so that
$E_c(T) \sim B_{\delta}(T,T_{\rm max})$.  If $T > T_{\rm max}$, then
equation (\AVEBT b) states that $E_c(T) = k T^{-1}$ where $k$ is a
constant.  In fact, when $T > T_{\rm max}$,  $E_c(T)$ is precisely 
a constant times $T^{-1}$
for all functions $P(T,T_e)$ and $E_c(T,T_e)$, regardless of their
functional form.  Even if $P(T,T_e)$ has a form that allows $T_{\rm
max}$ to formally go to infinity,
there must be an effective $T_{\rm max}$ where $P(T,T_e)$ 
approaches zero.  Otherwise, $V(T)$ would diverge or GRB profiles
would not have fast rises. Thus, in all cases, $E_c(T)$ follows
$T^{-1}$ in the decay phase. 

{}From a physical perspective, this universal spectral evolution can be
understood on the basis of symmetry.  We are not claiming that the
intrinsic variations in $P$ and $E_c$ are not manifested in the
observations. Indeed, while $P$ and $E_c$ are changing, the
observations track $P$ and $E_c$ (between $T
= 80$ and $T \sim 100$ in Fig.~2).  The universal spectral evolution only occurs in the
decay phase.
  During the decay phase, photons are delayed by the curvature of the
shell, so they come from some off-axis annulus at an angle $\theta =
\theta_1$.   By symmetry, the same averaging that
occurs at $\theta \sim 0$ also occurs at $\theta = \theta_1$; therefore,
the two spectra are identical, except that the average spectrum from 
$\theta = \theta_1$ is Lorentz
shifted by the boost $\Gamma(1-\beta\cos \theta_1)$ rather than
$\Gamma(1-\beta)$.  Equation (\BOOST) gives a one-to-one relationship of
the boost to the arrival time.  It is that boost that gives the
universal spectral evolution.

The disagreement between the time history evolution and the spectral
evolution
is further evidence that something must break the
local spherical symmetry if single relativistic shells are involved in GRBs.
For example, if the shell can act as a parallel slab or if the GRB
producing regions are confined to narrow pencil beams, then the
envelope and spectra can evolve uncoupled.
 In Fenimore, Madras, \& Nayakshin (1996),
we presented several other arguments that implied that LSS must be
broken or GRBs must be central engines.  For
example, precursors and the fact that many long complex 
bursts do not follow the
expected envelope imply that many parts of the shell 
never emit photons, resulting in a low efficiency for converting the 
kinetic energy of the shell
into gamma-rays. Sari \& Piran (1997) argue that the efficiency is
too low ($\sim 1$\%) to be consistent with merging  neutron stars and, thus,
one must resort to internal shock models (however, see
Fenimore 1997a).

In this paper, we show the need to break the LSS is also supported by
the spectral evolution.
 We have shown that intrinsic variations
alone are not sufficient to
explain  the observed
 spectral evolution.  This reinforces the ``shell symmetry'' problem we
introduced in Fenimore, Madras, and Nayakshin: GRB models that
involve a single relativistic shell
must explain how the material
is confined to pencil beams narrower than $\Gamma^{-1}$, how 
LSS is broken,  or how the
shell can be very inefficient without excessive energy requirements.

  In this paper, we did not vary $\Gamma$ with
 time.
Varying $\Gamma$ would not introduce any additional intrinsic
 spectral evolution since it would
 be equivalent to varying $E_c^{\prime}$, which is already accounted for
 in equation (\EC).  However, in a future paper we will show how
 changes in $\Gamma$ in time can affect the shape of the envelope and, perhaps,
 result in the breaking of LSS.

If the ``shell symmetry'' problem cannot be solved for the external
shock models, those models might have to be abandoned in favor of the
internal shock models, although these might have similar
problems.
In the internal shock models, peaks in the GRB time history arise from
variations in the central activity. 
We have fit equations (\ENVELOPE) and (\AVEBT) to individual
peaks in GRBs and the spectra evolve faster than $(T/T_0)^{-1}$. 
In general, characteristic energies in peaks can
evolve by factors of 4 whereas spectra from relativistic shells 
should only evolve by a factor $\sim 2$ across a peak.

If the internal shocks occur because fast shells run into slow shells
(Rees and \Mesz 1994) resulting in a shell with curvature, 
then the internal shock
models will also have to solve the ``shell symmetry'' problem. 
 The internal shock models also have to explain how a source
can produce $(\Omega/4\pi)10^{51}$ erg sec $^{-1}$ of activity for
hundreds of seconds. (Here, $\Omega/4\pi$ is the fraction of the sky
into which the burst beams its shells.)

\section*{References}

\BB
Band, D.~L., et al. 1993, ApJ, 413, 281





\BB
{Fenimore, E.~E., Madras, C.~D., \& Nayakshin, S., 1996, ApJ 473, 998,
astro-ph 9607163}

\BB
Fenimore, E.~E., 1997a, Proc. of 18th Texas Symposium on Relativistic
Astrophysics, eds. A.~Olinto, J.~Frieman, \& D.~Schram, astro-ph/9705028.

\BB
{Fenimore, E.~E., 1997b, ApJ,  to be submitted}



\BB
Klebesadel, R.~W., Strong, I., \& Olsen, R., ApJ, 1973, 182 L85


\BB
Katz J. I., 1994, ApJ, 422, 248



\BB
Liang, E., \& Kargatis 1996 Nature 381, 49

\BB
Lamb D. Q., 1995, PASP 107, 1152

\BB
Meegan C. A., et al., 1992, Nature, 335, 143


\BB
\Mesz P., \& Rees M. J., 1993, ApJ, 405, 278






\BB
\Pacz B., 1995, PASP 107, 1167



\BB
Piran T., Shemi A. \& Narayan R., 1993 MNRAS 263, 861


\BB
Rees M. J. \& \Mesz P., 1994, ApJ, 430, L93

\BB
Sari, R., \& Piran, T., 1997, ApJ, in press







\label{last}

\end{document}